# On The Leptons Masses

I. V. Grossu [(1)], C. Besliu [(1)]

[(1)] *University of Bucharest, Faculty of Physics, Bucharest-Magurele, Romania*

*In this paper we present a connection between the results obtained from a semiclassical study of a relativistic, two-body system, and the leptons masses. Some possible consequences are also discussed.*

Starting from the observation that the leptons masses could be approximated (Tab.1) using the empirical relation (1), and inspired by the atomic Bohr model [1], we tried to find some connections with the results obtained from the semiclassical study of a relativistic two-body problem.

$$m \cong \alpha n^\beta = m_e n^{7.5}, \quad n \in \{1,2,3\} \qquad (1)$$

where "$m_e$" is the electron mass.
A specific interaction was chosen:

$$F = a \cdot r^b \qquad (2)$$

where "$F$" represents the force, "$r$" the distance between the two particles, and "$a$", "$b$" are constants.

Working in the reference system related to one of the two constituents, we considered, for simplicity, the approximation of the circular uniform motion. By applying stationary de Broglie waves [2], and the centripetal force (2):

$$\begin{cases} mvr = n\hbar \\ \\ \dfrac{mv^2}{r} = ar^b \end{cases} \qquad (3)$$

we obtained the mass spectrum, and the radius:

$$m_n = \frac{a}{v^2} \left( \frac{\hbar v}{a} \right)^{\frac{b+1}{b+2}} n^{\frac{b+1}{b+2}} \qquad (4)$$

$$r_n = \left(\frac{\hbar v}{a}\right)^{\frac{1}{b+2}} n^{\frac{1}{b+2}} \quad (5)$$

where "$m$" is the movement mass, "$r$" the radius, "$n$" the principal quantum number, "$v$" the velocity, and "$h$" the Planck constant.

One can observe that, if the rest mass tends to zero, and the velocity tends to the speed of light "$c$", the equation (4) has the same form as the empirical relation (1):

$$m_n = \frac{a}{c^2}\left(\frac{\hbar c}{a}\right)^{\frac{b+1}{b+2}} n^{\frac{b+1}{b+2}} \Rightarrow \frac{b+1}{b+2} = 7.5 \quad (6)$$

$$r_n = r_1 n^{-6.5} \quad (7)$$

| Particle | Lepton mass/Electron mass [3] | n | Lepton mass/Electron mass (1) |
|---|---|---|---|
| e | 1 | 1 | 1 |
| μ | 207 | 2 | 181 |
| τ | 3484 | 3 | 3787 |
| Searches | >196078 | 4 | 32768 |

Tab.1 Relative agreement between the empirical relation (1) and the known values for leptons masses [3].

## CONCLUSIONS

Based on an empirical observation (1), we applied a semiclassical study for a relativistic system composed by two identical particles with negligible rest mass. Considering a specific interaction type (2), for the approximate case of the circular uniform motion, we obtained some encouraging results:
- a relative agreement (Tab.1) between the energy spectrum (6) and the leptons masses [3];
- by applying the Lorenz contraction, the system radius tends to zero when the velocity tends to the speed of light. This fact is in a good agreement with the leptons properties (considered point-likes).

Some predictions can also be considered:
- the existence of a structure corresponding to e, μ, and τ leptons;
- the mass for a possible fourth lepton (Tab.1).
- the system radius decreases rapidly with the principal quantum number (7). This fact could be intuitively related with a limitation for the number of leptons families.

## REFERENCES

[1] Niels Bohr (1914). "The spectra of helium and hydrogen". Nature 92: 231-232.
[2] L. de Broglie, Recherches sur la théorie des quanta (Researches on the quantum theory), Thesis (Paris), 1924; L. de Broglie, Ann. Phys. (Paris) 3, 22 (1925). Reprinted in Ann. Found. Louis de Broglie 17 (1992) p.
[3] W.-M. Yao et al. (Particle Data Group), J. Phys. G 33, 1 (2006) (URL: http://pdg.lbl.gov)